\documentclass[12pt]{article}
\usepackage{amsmath}
\usepackage{amssymb}
\usepackage{amsfonts}
\usepackage{latexsym}
\usepackage{color}
\usepackage{graphicx}

\catcode `\@=11 \@addtoreset{equation}{section}

\catcode `\@=12

\newcommand{\be}{\begin{equation}}
\newcommand{\en}{\end{equation}}
\newcommand{\bea}{\begin{eqnarray}}
\newcommand{\ena}{\end{eqnarray}}
\newcommand{\beano}{\begin{eqnarray*}}
\newcommand{\enano}{\end{eqnarray*}}
\newcommand{\bee}{\begin{enumerate}}
\newcommand{\ene}{\end{enumerate}}

\newcommand{\Hil}{{\cal H}}

\newcommand{\F}{{\cal F}}

\newcommand{\Sc}{{\cal S}}
\newcommand{\Rc}{{\cal R}}

\newcommand{\1}{1 \!\! 1}
\newtheorem{thm}{Theorem}

\newtheorem{defn}[thm]{Definition}

\begin{document}
\thispagestyle{empty}

\vspace*{1cm}

\begin{center}
{\Large \bf One-directional quantum mechanical dynamics and an application to decision making}   \vspace{2cm}\\

{\large F. Bagarello}
\vspace{3mm}\\[0pt]
Dipartimento di Ingegneria, Scuola Politecnica,\\[0pt]
Universit\`{a} di Palermo, I - 90128 Palermo,  and\\
\, INFN, Sezione di Napoli, Italy.\\[0pt]
E-mail: fabio.bagarello@unipa.it\\[0pt]
home page: www1.unipa.it/fabio.bagarello \vspace{8mm}\\[0pt]
\end{center}

\vspace*{0.5cm}

\begin{abstract}
	In recent works we have used quantum tools in the analysis of the time evolution of several macroscopic systems. The main ingredient in our approach is the self-adjoint Hamiltonian $H$ of the system $\Sc$. This Hamiltonian quite often, and in particular for systems with a finite number of degrees of freedom, gives rise to reversible and oscillatory dynamics. Sometimes this is not what physical reasons suggest. We discuss here how to use non self-adjoint Hamiltonians to overcome this difficulty: the time evolution we obtain out of them show a preferable arrow of time, and it is not reversible. Several applications are constructed, in particular in connection to information dynamics. 

\end{abstract}

\vspace{2cm}


\vfill

\newpage

\section{Introduction}\label{sect1}

In a series of papers and books quantum mechanical tools and ideas have been used in connection with systems which are not apparently connected with quantum mechanics at all. For instance, biology, economy, psychology, are realms of the research which have been considered, and still are considered, using this strategy. We refer to \cite{qal}-\cite{paulina} for some recent books and few recent papers, where several other references can be found. 

Some of the applications considered in the literature are related to the dynamics of certain macroscopic systems. And, in many applications, the interest is focused on the asymptotic limit of some function describing the time evolution of the system under analysis. This is, for instance, often the case in Decision Making, where people are generally interested in the derivation of the final decision of the agents of the system. For instance, in \cite{bag5} and \cite{Bagarello2015b}, we have analyzed the long time behaviour of a love story between two lovers and, as a rather different application, what has been called the {\em decision function} of three political parties, describing their attitude to form, or not, some political alliance. Again, our main interest was on the asymptotic limits of these decision functions. The model was refined in \cite{baggarg}, while more applications of the same kind are described in \cite{baghavkhr} and \cite{baggame}. The main technique used in all these papers is based on the use of some self-adjoint Hamiltonian $H$ constructed in terms of suitable ladder operators, like creation and annihilation operators satisfying canonical (anti-) commutation relations. $H$ is used in order to deduce the time evolution of some {\em observables} of the system $\Sc$. These observables are self-adjoint (number) operators with some relevance for the description of $\Sc$. This strategy has been used several times along the years, in many different contexts, some of them described in detail in \cite{bagbook} and \cite{bagbook2}. The limitation of this strategy is the following: using a standard Heisenberg dynamics, or its dual Schr\"odinger counterpart, the dynamics we can get is quite often periodic or quasi-periodic, so that, but for trivial situations in which the system does not evolve at all, no asymptotic limit can be found. This can be, in fact, rigorously derived if $\Sc$ lives in a finite dimensional Hilbert space and $H=H^\dagger$. If the Hilbert space is infinite dimensional, then this is not so obvious, but it happens quite often for several systems. This suggests that our quantum-like strategy should be somehow enriched, if we are interested in describing systems for which some equilibrium is expected, after some time. And in fact, several proposals have been considered along the years to achieve this aim. An equilibrium for  $\Sc$ can be found, for instance, if $\Sc$ is {\em open}, i.e., if it interacts with some external infinite-dimensional reservoir. This is the approach proposed, for instance, in \cite{bag5}-\cite{baggame}. In this case the Hamiltonian of the system is self-adjoint, $H=H^\dagger$, but it includes, in particular, the interaction between $\Sc$ and its reservoir (the environment) $\Rc$. This approach is interesting, and physically motivated, since in all the examples considered so far $\Rc$ is not just a mathematical trick to obtain some stable asymptotic behaviour of the observables of $\Sc$, but it has an explicit meaning relevant for the model. 

A completely different approach, still producing some equilibrium for large time, was introduced in \cite{therule} and then analyzed in a series of papers, \cite{HRO2,rosa1,rosa2}. In this case, the system $\Sc$ is again described by a self-adjoint Hamiltonian $H$, but $\Sc$ does not interact with any reservoir. However, periodically, $\Sc$ is subjected to some sort of (external or internal) check which can slightly modify some of the aspects of $\Sc$, depending on the output of the check. This is what we have called $(H,\rho)$--induced dynamics, to stress the fact that the time evolution of $\Sc$ is driven by both $H$ and by the check (the {\em rule}) $\rho$. For instance, in \cite{therule} the rule $\rho$ was used to propose a quantum-like version of the game of life. At fixed time intervals, $\rho$ checked the densities of the populations in the cells surrounding a given cell $C_0$ in the lattice where $\Sc$ lives. The values of these densities decide if, in the next generation, $C_0$ is dead or alive. We have seen in \cite{HRO2}  how the presence of $\rho$ in the analysis of the time evolution of some systems can produce an asymptotic limit for their observables. 

Another possibility to get an asymptotic limit for $\Sc$ has also been considered in the past years. The idea is to replace some of the real parameters of the Hamiltonian $H$ of $\Sc$ with other, complex valued ones. The signs of these complex parameters are connected with their meaning, and with the effective result of their presence in $H$. For instance, in \cite{BCO2016}, a negative imaginary part of some parameter of the (free) Hamiltonian $H_0$ of the system (soil+seeds+plants) was used to model the presence of stress factors, while a positive imaginary part of other parameters of $H_0$ was used to model some positive effect acting continuously on the system. Playing with the parameters, we were able to produce an equilibrium for the system {\em far from desertification}. A similar idea was also used in the description of closed ecosystems, \cite{BagOli2014}, to model systems whose efficiency in recycling garbage into nutrients is not perfect. It is clear that, in this way, we are giving up the hypothesis that the full Hamiltonian of the system should be self-adjoint. However, the kind of non-hermitianity that we have considered in \cite{BCO2016} and \cite{BagOli2014} is of a very special kind: if the parameters used in the definition of $H$ are taken to be real, then $H=H^\dagger$.

More recently, \cite{baggarg2}, we have considered a different choice of non self-adjointness for $H$, in the biological context of cells proliferation. The Hamiltonian ceases to be self-adjoint not because of the presence of some complex parameters, but because some operator in $H$ is not paired to its adjoint counterpart: if $A$ and $B$ are (ladder) operators used in the analysis of the cells, $H$ contains terms like $A^\dagger B$, but not its adjoint, $B^\dagger A$. This makes of $H$ a non self-adjoint Hamiltonian, with all the problems, and the possibilities, that this choice produces. In particular, the time evolution appears to be {\em one-directional}: some fluxes seem more preferable than  others. In particular, if we have no medical treatment acting on the cells, it is natural to expect that the healthy cells become sick, but not viceversa. This effect was well described with our choice of $H$ in \cite{baggarg2}, which is the first time, in our knowledge, in which the time evolution of a biological system is given in terms of such an Hamiltonian operator. Something along the same lines, but in an economical context, can be found in \cite{havkhrebook}, where, however, ladder operators play no role. Creation and annihilation operators appear again, in constructing non self-adjoint Hamiltonians used in Finance, in \cite{KHR} and in \cite{bagBS}. Going back to biology, our conclusions in \cite{baggarg2} suggest that non self-adjoint Hamiltonians of the kind considered there works well, and this would open many possible lines of research in the future. However, before considering more complicated (and useful) applications, we prefer to study in details what happens when our system $\Sc$ is attached to some special non self-adjoint Hamiltonian, to understand the basic mechanisms which then we can try to adapt in the analysis of more complicated and more realistic systems. This is exactly what we will discuss in some details in this paper, which is organized as follows: in the next section we begin our analysis and we discuss different approaches, discussing their pros and cons at a general level. In Section \ref{sect3} we will show that only one, among all the possibilities, is in agreement with what one could expect. In Section \ref{sect4} we use this particular choice to model a simple system of information dynamics. The model is simple enough to allow for an almost entirely analytical treatment, which makes it possible to understand well the details of our framework. Section \ref{sect5} contains our conclusions.

\section{Comparing strategies: theory}\label{sect2}

In the first part of this section we will briefly describe  the dynamics of a quantum-like system when this is described by a self-adjoint Hamiltonian $H_0$, $H_0=H_0^\dagger$, and, to keep the treatment simple, time independent. This information can be found in any book in quantum mechanics, see \cite{mer,mess} for instance, or, more in line with our particular situation, in \cite{bagbook,bagbook2}. In view of our particular interest in this paper, we will also assume that the system $\Sc$ is {\em closed}. At $t=0$ $\Sc$ is described by a wave function $\Psi(0)$, whose time evolution satisfies the Schr\"odinger equation $i\dot{\Psi}(t)=H_0\Psi(t)$. Its (formal) solution is $\Psi(t)=e^{-iH_0t}\Psi(0)$. The reason why this solution is {\em formal} is because, in general, computing $e^{-iH_0t}$ is an hard technical problem, mainly if $\Sc$ lives in an infinite-dimensional Hilbert space $\Hil$. On the other hand, if $dim(\Hil)$ is small enough, this computation could be easy or, at least, reasonably simple. We call {\em observable} of the system every self-adjoint operator $X$ which correspond to some measurable quantity of $\Sc$. Following our general strategy,  see \cite{bagbook,bagbook2}, our main aim is to compute the mean value $x(t)=\left<\Psi(t),X\Psi(t)\right>$ or, equivalently,
\be
x(t)=\left<\Psi(t),X\Psi(t)\right>=\left<\Psi(0),X(t)\Psi(0)\right>, \qquad\mbox{where }\, X(t)=e^{iH_0t}Xe^{-iH_0t}.
\label{21}\en
Notice that $\dot{X}(t)=i[H_0,X(t)]$, which is known as the Heisenberg equation of motion for $X(t)$.
The two mean values in (\ref{21}) are respectively connected to the so-called Schr\"odinger and Heisenberg representations, depending on the fact that we are {\em attaching} the time dependence of the system to the state $\Psi$, as in $\Psi(t)=e^{-iH_0t}\Psi(0)$ or to the observable $X$, as in $X(t)=e^{iH_0t}Xe^{-iH_0t}$. The equality  in (\ref{21}) is a simple consequence of the unitarity of the operator $e^{-iH_0t}$, which follows from the self-adjointness of $H_0$. 

What is interesting for us is to see what happens if we consider an Hamiltonian which is no longer self-adjoint. In other words, we assume that the system $\Sc$, described at $t=0$ by a normalized vector $\Psi(0)$ in the Hilbert space $\Hil$, has a dynamics which is driven by a non self-adjoint Hamiltonian $H$: $H\neq H^\dagger$. Our particular interest is in finding the analogous of $x(t)$ in (\ref{21}), for some observable $X$. Incidentally we can safely assume that, for $t=0$, $X=X^\dagger$. This relation could be maintained or not during the time evolution. This depends on how we describe the evolution of $\Sc$, using the Schr\"odinger or the Heisenberg representation. The main output of our analysis will be that there is only one possibility which gives rise to solutions which are in agreement with the general strategies proposed and analyzed in \cite{bagbook,bagbook2}, if $H\neq H^\dagger$. Here we only mention how the different possibilities look like, while we will see later which one among these should be chosen. This choice will be shown to be in agreement with the one in \cite{baggarg2}, and will be used to deduce more results on a simple model for information dynamics.

\begin{enumerate}

\item Our first choice is a {\em minimal displacement} from the one in (\ref{21}). We assume that the wave function for $\Sc$, $\Psi(t)$,
 evolves according to the same Schr\"odinger equation as if $H$ were self-adjoint. Hence, calling $\Psi(0)$ the initial value of $\Psi(t)$, we have
 \be
 i\dot{\Psi}(t)=H\Psi(t), \qquad \mbox{ and }\quad \Psi(t)=U(t)\Psi(0),
 \label{22}\en
where $U(t)=e^{-iHt}$. Then we can introduce the following function:
\be
x^{(1)}(t)=\left<\Psi(t),X\Psi(t)\right>.
\label{23}\en
Here the suffix $(1)$ stands for 'first choice'. We stress that the essential difference with respect to (\ref{21}) is that $U(t)$ is not an unitary operator. In fact, $U^{-1}(t)=e^{iHt}$, while $U^\dagger(t)=e^{iH^\dagger t}$. Hence, if we want to extend formula (\ref{21}), it is clear that we have
\be
x^{(1)}(t)=\left<\Psi(t),X\Psi(t)\right>=\left<\Psi(0),X^{(1)}(t)\Psi(0)\right>,
\label{24}\en
where $X^{(1)}(t)=e^{iH^\dagger t}Xe^{-iH t}$, which would coincide with its standard Heisenberg dynamics if $H=H^\dagger$, but not in general.

\item If we insist with the choice of $X^{(1)}(t)$ as the tentative time evolution of $X$ in the present situation, we immediately see that this time evolution is not an automorphism:  given two observables of $\Sc$, $X$ and $Y$, we have in general $(XY)^{(1)}(t)\neq X^{(1)}(t)Y^{(1)}(t)$. This creates, among other problems, several technical difficulties. In particular, it is quite hard in concrete, and even simple systems, to produce a closed set of differential equations which describe the dynamics. For this reason, it makes sense to consider a dual approach with respect to that in (\ref{22})-\eqref{24}: rather that taking the Schr\"odinger equation as the starting point of our procedure, we take now the standard Heisenberg equation for $X$ as the first step. In other words, we assume that the time evolution of the observable $X$, which we now call $X^{(2)}(t)$, obeys the equation
\be
\frac{dX^{(2)}(t)}{dt}=i[H,X^{(2)}(t)],
\label{25}\en
even if $H\neq H^\dagger$. This equation can be easily solved (again, formally), and the result is
\be
X^{(2)}(t)=e^{iHt}Xe^{-iHt}.
\label{26}\en
It is clear that $(XY)^{(2)}(t)= X^{(2)}(t)Y^{(2)}(t)$, but is is also clear that
$$
\left<\Psi(0),X^{(2)}(t)\Psi(0)\right>\neq \left<\Psi(t),X\Psi(t)\right>,
$$
where, as usual, $\Psi(t)=U(t)\Psi(0)$. With this choice the relevant quantity to compute is
\be
x^{(2)}(t)=\left<\Psi(0),X^{(2)}(t)\Psi(0)\right>,
\label{27}\en
which replace $x^{(1)}(t)$ above. Therefore, $x^{(1)}(t)\neq x^{(2)}(t)$, in general.

\item A third possibility arises from a simple consideration: when $U(t)$ is not unitary, the norm of $\Psi(t)$ is not preserved, in general. This is a problem which is widely discussed in some literature on PT-quantum mechanics, for instance, where the probabilistic interpretation of the wave function is recovered only when $\Psi(t)$ is replaced by $\frac{\Psi(t)}{\|\Psi(t)\|}$. We refer to \cite{sgh,bagaop2015} for many considerations on this aspect for quantum mechanical systems. Then we introduce $x^{(3)}(t)$ as a simple normalized version of (\ref{23}):
\be
x^{(3)}(t)=\left<\hat\Psi(t),X\hat\Psi(t)\right>, \quad \mbox{ where } \hat\Psi(t)=\frac{\Psi(t)}{\|\Psi(t)\|}=\frac{U(t)\Psi(0)}{\|U(t)\Psi(0)\|}.
\label{28}\en
Of course, there is no analogous natural way to introduce a normalized version of $x^{(2)}(t)$, even if we could think to replace $X^{(2)}(t)$ with 
$$
\frac{X^{(2)}(t)}{\|U(t)\|\|U(t)^\dagger\|}.
$$
However, in many relevant cases, $H$ is unbounded  and, being non self-adjoint, it would give rise to an operator $U(t)$ which, most likely, is unbounded as well. Hence the denominator in this fraction diverges for some (or many) $t$,  and the definition would make no sense. This is not the case in (\ref{28}), since $\Psi(0)$ belongs to the domain of $U(t)$ for all $t$.  

\end{enumerate}

In the next section we will compare these three strategies and, based on the role and the meaning of the ladder operators in concrete systems, we will see that the correct choice (for us) is the one in (\ref{28}). To clarify this point, we will consider three simple systems with a finite (and small) number of degrees of freedom, with two or three agents and obeying different commutation relations, to cover different situations, so to make our conclusions more robust.

\section{Comparing strategies: applications}\label{sect3}

In this section we will consider three different models and discuss for them $x^{(j)}(t)$, $j=1,2,3$, to understand which is the correct way to introduce a time evolution for systems driven by non self-adjoint Hamiltonians. As already stated, we will conclude that formula (\ref{28}) gives the correct recipe to use, in this situation.

\subsection{Model 1: two agents and two levels}

The first model we want to describe is defined by a manifestly non self-adjoint Hamiltonian $H=\lambda a_2^\dagger a_1$, where $\lambda>0$ is the coupling constant\footnote{Working with positive $\lambda$ is useful just to fix the details. We do not expect serious changes for $\lambda<0$.} between agents $\tau_1$ and $\tau_2$, which define the system $\Sc$. $a_1$ and $a_2^\dagger$ are ladder operators for $\tau_1$ and $\tau_2$ respectively, obeying the following canonical anti-commutation relations:
\be
\{a_k,a_j^\dagger\}=\delta_{k,j}\1, \qquad \qquad a_j^2=0,
\label{31}\en
$j,k=1,2$, and $\1$ is the identity operator on the Hilbert space of the system, $\Hil=\mathbb{C}^4$. Following our standard approach, \cite{bagbook,bagbook2}, we introduce an orthonormal basis of $\Hil$ as follow: $\F_\varphi=\{\varphi_{kj}, \, k,j=0,1\}$, where
$$
a_1\varphi_{00}=a_2\varphi_{00}=0, \qquad \varphi_{10}=a_1^\dagger\varphi_{00}, \qquad \varphi_{01}=a_2^\dagger\varphi_{00}, \qquad \varphi_{11}=a_1^\dagger a_2^\dagger\varphi_{00}.
$$
An explicit expression for these operators and vectors is easily found:
$$
a_1=\left(
\begin{array}{cccc}
0 & 1 &  0 & 0 \\
0 & 0 & 0 & 0  \\
0 & 0 & 0 & 1 \\
0 & 0 & 0 & 0 \\
\end{array}
\right),\qquad a_2=\left(
\begin{array}{cccc}
0 & 0 &  1 & 0 \\
0 & 0 & 0 & -1  \\
0 & 0 & 0 & 0 \\
0 & 0 & 0 & 0 \\
\end{array}
\right), $$
and $$\varphi_{00}=\left(
\begin{array}{c}
1 \\
0   \\
0  \\
0  \\
\end{array}
\right), \quad \varphi_{10}=\left(
\begin{array}{c}
0 \\
1   \\
0  \\
0  \\
\end{array}
\right), \quad \varphi_{01}=\left(
\begin{array}{c}
0 \\
0   \\
1  \\
0  \\
\end{array}
\right), \quad \varphi_{11}=\left(
\begin{array}{c}
0 \\
0   \\
0  \\
1  \\
\end{array}
\right).
$$
Hence
$$
H=\lambda\left(
\begin{array}{cccc}
0 & 0 &  0 & 0 \\
0 & 0 & 0 & 0  \\
0 & 1 & 0 & 0 \\
0 & 0 & 0 & 0 \\
\end{array}
\right), \quad \mbox{ and }\quad  U(t)=e^{-iHt}=\left(
\begin{array}{cccc}
1 & 0 &  0 & 0 \\
0 & 1 & 0 & 0  \\
0 & -it\lambda & 1 & 0 \\
0 & 0 & 0 & 1 \\
\end{array}
\right).
$$
We see that both $H$ and $U(t)$ have very simple expressions, which also makes clear the fact that $H\neq H^\dagger$ and that $U(t)$ is not unitary. The observables of the system are the number operators $N_j=a_j^\dagger a_j$, $j=1,2$. They are both diagonal matrices with $N_1=diag(0,1,0,1)$ and $N_2=diag(0,0,1,1)$. What we expect from our $H$ is that it increases the eigenvalue of $N_2$ (because of $a_2^\dagger$) and simultaneously decreases the eigenvalue of $N_1$ (because of $a_1$). And, more important, we expect this process is irreversible, since the Hamiltonian does not contain the adjoint term $a_1^\dagger a_2$. We begin our analysis computing the time evolution of the mean values of $N_j$ as in  (\ref{23}), choosing first $\Psi(0)=\varphi_{10}$. This is because this state corresponds to an initial value of eigenvalues of $N_1$ and $N_2$ equal to 1 and 0, which is the only case in which the first number can decrease and the second can increase\footnote{Recall that fermionic number operators, like $N_1$ and $N_2$, can only have 0 and 1 as eigenvalues.}. Simple computations show that
$$
n_1^{(1)}(t)=\left<\Psi(t),N_1\Psi(t)\right>=1, \qquad n_2^{(1)}(t)=\left<\Psi(t),N_2\Psi(t)\right>=(\lambda t)^2.
$$
This is not what we expect, for two reasons: (i) $n_1^{(1)}(t)$ does not decrease, as expected because of the presence of $a_1$ in $H$, and (ii) $n_2^{(1)}(t)$ increases above the maximum allowed (we should always have $n_j(t)\in[0,1]$, for all realistic expression for $n_j(t)$). Henceforth, the choice $x^{(1)}(t)$ for the dynamics of a system driven by a non self-adjoint Hamiltonian does not work already for this quite simple model, and will not be considered further in this paper.

Let us now check if the second choice works better. For that, it is convenient to observe that $N_j^{(2)}(t)$, $j=1,2$, obey the following standard Heisenberg equations of motion, $$ \frac{d N_j^{(2)}(t)}{dt}=ie^{iHt}[H,N_j]e^{-iHt}=i[H,N_j^{(2)}(t)], $$
with the initial condition $N_j^{(2)}(0)=N_j$, $j=1,2$. Now, since $[H,N_1]=H$ and $[H,N_2]=-H$, we conclude that $N_1^{(2)}(t)=N_1+iHt$ and $N_2^{(2)}(t)=N_2-iHt$. Hence, recalling that $\Psi(0)=\varphi_{10}$,
$$
n_1^{(2)}(t)=\left<\Psi(0),N_1^{(2)}(t)\Psi(0)\right>=1, \qquad n_2^{(2)}(t)=\left<\Psi(0),N_2^{(2)}(t)\Psi(0)\right>=0.
$$
This case is, in a sense, even worse than the previous one: both mean values stay constant, as if there was no dynamics at all.

We are left with the third possibility, the one in (\ref{28}). We see that $x^{(3)}(t)$ is related to $x^{(1)}(t)$ by a simple time-dependent normalization factor:
$$
x^{(3)}(t)=\frac{x^{(1)}(t)}{\|\Psi(t)\|^2}.
$$
Then, what we have to do is to compute $\|\Psi(t)\|$. We have
$$
\Psi(t)=U(t)\varphi_{10}=\left(
\begin{array}{cccc}
1 & 0 &  0 & 0 \\
0 & 1 & 0 & 0  \\
0 & -it\lambda & 1 & 0 \\
0 & 0 & 0 & 1 \\
\end{array}
\right)\left(
\begin{array}{c}
0 \\
1   \\
0  \\
0  \\
\end{array}
\right)=\left(
\begin{array}{c}
0 \\
1   \\
-it\lambda  \\
0  \\
\end{array}
\right), 
$$
so that $\|\Psi(t)\|^2=1+(\lambda t)^2$. Hence we find
$$
n_1^{(3)}(t)=\frac{n_1^{(1)}(t)}{1+(\lambda t)^2}=\frac{1}{1+(\lambda t)^2}, \qquad n_2^{(3)}(t)=\frac{n_2^{(1)}(t)}{1+(\lambda t)^2}=\frac{(\lambda t)^2}{1+(\lambda t)^2}.
$$
We see that these functions behave exactly as we  expect: $n_1^{(3)}(t)$ decrease from its original value, 1, to the value 0. This is the effect of $a_1$ in $H$. Also, $n_2^{(3)}(t)$ increases from 0 to 1, as an effect of the presence of $a_2^\dagger$ in $H$. Hence the conclusion of this preliminary example is the following: if we want to describe a one-directional time evolution using some $H\neq H^\dagger$, the only possibility which seems to work so far is the one in (\ref{28}). From now on we will focus our attention on this formula, and remove the suffix $(3)$ since the other two will not be considered any more in this paper: they simply do not work! Hence we will restrict to the following definition:

\begin{defn}\label{def1}
	Let $\Sc$ be a system driven by an Hamiltonian $H$, not necessarily self-adjoint. Suppose that, at $t=0$, $\Sc$ is in the normalized state $\Psi(0)$. Let $X$ be an observable for $\Sc$, i.e. a self-adjoint operator relevant for the description of $\Sc$. Hence the classical dynamics of the operator $X$ on the initial state $\Psi(0)$ is \be
	x(t)=\left<\hat\Psi(t),X\hat\Psi(t)\right>, \quad \mbox{ where } \hat\Psi(t)=\frac{\Psi(t)}{\|\Psi(t)\|}=\frac{U(t)\Psi(0)}{\|U(t)\Psi(0)\|}.
	\label{32}\en
\end{defn}

It is clear that, if $H=H^\dagger$, $\|\Psi(t)\|=\|\Psi(0)\|=1$, and we go back to (\ref{21}). Otherwise we get something different. It is also clear that there is no immediate Heisenberg counterpart for (\ref{32}).

Needless to say, the analysis above suggests that Definition \ref{def1} is reasonable, but it is far from proving that it is the correct one, for our systems. For this reason, to make the definition more robust, we will now check what happens with different choices of $\Psi(0)$ and then, in the next sections, we will use (\ref{32}) in the analysis of other, more complicated systems.

For the present system it is quite easy to check that $U(t)\varphi_{jk}=\varphi_{jk}$, if $jk\neq10$. Then, if $\Psi(0)=\varphi_{jk}$, $jk\neq10$, $\Psi(t)=\Psi(0)$, $\|\Psi(t)\|=1$ and from (\ref{32}) we conclude that $n_j(t)=n_j(0)$. In other words: the Hamiltonian considered here is only able to modify $\Sc$ when this is in the state $\varphi_{10}$. Otherwise, it leaves the system unchanged. This is due to the fact that, since $H^2=H^3=\ldots=0$,
$$
U(t)=e^{-iHt}=\1-iHt,
$$
and that $H$ destroys all the states different from $\varphi_{10}$.

\subsection{Model 2: two agents and three levels}

The model we consider here has the same number of agents, $\tau_1$ and $\tau_2$ and the same formal Hamiltonian, which we call again $H$: $H=\lambda A_2^\dagger A_1$, with some $\lambda>0$. The difference is in the nature of $A_1$ and $A_2$, which are no longer assumed here to be fermionic operators. The rationale for this is that, with this choice, we will show that our Definition \ref{def1} works well also for other kind of ladder operators, at least in this situation with two agents. This suggests that our results are not linked to the commutation relations assumed for the model.

The operators $A_j$ are constructed as in \cite{baggarg2}. We have
$$
A_1=\left(
\begin{array}{ccccccccc}
0 & 1 &  0 & 0 & 0 & 0 & 0 & 0 & 0\\
0 & 0 &  \sqrt{2} & 0 & 0 & 0 & 0 & 0 & 0\\
0 & 0 &  0 & 0 & 0 & 0 & 0 & 0 & 0\\
0 & 0 &  0 & 0 & 1 & 0 & 0 & 0 & 0\\
0 & 0 &  0 & 0 & 0 & \sqrt{2} & 0 & 0 & 0\\
0 & 0 &  0 & 0 & 0 & 0 & 0 & 0 & 0\\
0 & 0 &  0 & 0 & 0 & 0 & 0 & 1 & 0\\
0 & 0 &  0 & 0 & 0 & 0 & 0 & 0 & \sqrt{2}\\
0 & 0 &  0 & 0 & 0 & 0 & 0 & 0 & 0\\
\end{array}
\right), \quad A_2=\left(
\begin{array}{ccccccccc}
0 & 0 &  0 & 1 & 0 & 0 & 0 & 0 & 0\\
0 & 0 &  0 & 0 & 1 & 0 & 0 & 0 & 0\\
0 & 0 &  0 & 0 & 0 & 1 & 0 & 0 & 0\\
0 & 0 &  0 & 0 & 1 & 0 & \sqrt{2} & 0 & 0\\
0 & 0 &  0 & 0 & 0 & 0 & 0 & \sqrt{2} & 0\\
0 & 0 &  0 & 0 & 0 & 0 & 0 & 0 & \sqrt{2}\\
0 & 0 &  0 & 0 & 0 & 0 & 0 & 0 & 0\\
0 & 0 &  0 & 0 & 0 & 0 & 0 & 0 & 0\\
0 & 0 &  0 & 0 & 0 & 0 & 0 & 0 & 0\\
\end{array}
\right).
$$
These operators commute, $[A_1^\sharp,A_2^\sharp]=0$, where $A_j^\sharp$ is either $A_j$ or $A_j^\dagger$, and satisfy the equalities $A_1^3=A_2^3=0$. Moreover,
$$
[A_1,A_1^\dagger]=diag(1,1,-2,1,1,-2,1,1,-2), \qquad [A_2,A_2^\dagger]=diag(1,1,1,1,1,1,-2,-2,-2).
$$
We can use $A_1^\dagger$ and $A_2^\dagger$, together with the vacuum of $A_1$ and $A_2$, $\Phi_{00}=(1\,\, 0\,\, 0\,\, 0\,\, 0\,\, 0\,\, 0\,\, 0\,\, 0 )^T$, (here $T$ is the transpose), to construct an orthonormal basis $\F_\Phi$ of the Hilbert space for this model, $\Hil=\mathbb{C}^9$. The vectors of $\F_\Phi$ are constructed as follows:
$$
\Phi_{10}=A_1^\dagger \Phi_{00}, \quad \Phi_{01}=A_2^\dagger \Phi_{00}, \quad \Phi_{11}=A_1^\dagger A_2^\dagger \Phi_{00},\quad \Phi_{20}=\frac{1}{\sqrt{2}}{A_1^\dagger}^2 \Phi_{00}, \quad \Phi_{02}=\frac{1}{\sqrt{2}}{A_2^\dagger}^2 \Phi_{00}, $$

$$ \Phi_{21}=\frac{1}{\sqrt{2}}{A_1^\dagger}^2A_2^\dagger \Phi_{00}, \quad \Phi_{12}=\frac{1}{\sqrt{2}}A_1^\dagger{A_2^\dagger}^2 \Phi_{00},\quad \Phi_{22}=\frac{1}{2}{A_1^\dagger}^2{A_2^\dagger}^2 \Phi_{00}.
$$
Now, $N_1=A_1^\dagger A_1$ and $N_2=A_2^\dagger A_2$ are the following diagonal matrices: 
$$
N_1=diag(0,1,2,0,1,2,0,1,2), \qquad N_2=diag(0,0,0,1,1,1,2,2,2),
$$
while $H$ can be written as the following sparse matrix:
$$
H=\lambda\left(
\begin{array}{ccccccccc}
0 & 0 &  0 & 0 & 0 & 0 & 0 & 0 & 0\\
0 & 0 &  0 & 0 & 0 & 0 & 0 & 0 & 0\\
0 & 0 &  0 & 0 & 0 & 0 & 0 & 0 & 0\\
0 & 1 &  0 & 0 & 0 & 0 & 0 & 0 & 0\\
0 & 0 &  \sqrt{2} & 0 & 0 & 0 & 0 & 0 & 0\\
0 & 0 &  0 & 0 & 0 & 0 & 0 & 0 & 0\\
0 & 0 &  0 & 0 & \sqrt{2} & 0 & 0 & 0 & 0\\
0 & 0 &  0 & 0 & 0 & 2 & 0 & 0 & 0\\
0 & 0 &  0 & 0 & 0 & 0 & 0 & 0 & 0\\
\end{array}
\right),
$$
clearly non self-adjoint, which produces 
$$
U(t)=\left(
\begin{array}{ccccccccc}
1 & 0 &  0 & 0 & 0 & 0 & 0 & 0 & 0\\
0 & 1 &  0 & 0 & 0 & 0 & 0 & 0 & 0\\
0 & 0 &  1 & 0 & 0 & 0 & 0 & 0 & 0\\
0 & -i\lambda t &  0 & 1 & 0 & 0 & 0 & 0 & 0\\
0 & 0 &  -i\lambda\sqrt{2}\,t & 0 & 1 & 0 & 0 & 0 & 0\\
0 & 0 &  0 & 0 & 0 & 1 & 0 & 0 & 0\\
0 & 0 &  -\lambda^2 t^2 & 0 & -i\lambda \sqrt{2}\,t & 0 & 1 & 0 & 0\\
0 & 0 &  0 & 0 & 0 & -2i\lambda t & 0 & 1 & 0\\
0 & 0 &  0 & 0 & 0 & 0 & 0 & 0 & 1\\
\end{array}
\right)
$$
Let us now consider, as initial state, the vector $\Psi(0)=\Phi_{11}$. Using $U(t)$ as above, and formula (\ref{32}) we get
$$
n_1(t)=\left<\hat\Psi(t),N_1\hat\Psi(t)\right>=\frac{1}{1+2\lambda^2 t^2}, \qquad n_2(t)=\left<\hat\Psi(t),N_2\hat\Psi(t)\right>=\frac{1+4\lambda^2 t^2}{1+2\lambda^2 t^2}.
$$
Hence we see that $n_1(t)$ decreases from its original value, $n_1(0)=1$, to zero, while $n_2(t)$ increases from its original value, $n_2(0)=1$, to the maximum value allowed by our model, 2. This result is in full agreement with the expression of $H=\lambda A_2^\dagger A_1$. As in the previous model, $\lambda$ determines the speed of convergence of $n_j(t)$ to its asymptotic value: the higher its value, the faster the convergence.

Another interesting choice for $\Psi(0)$ is the following: $\Psi(0)=\Phi_{21}$. Repeating the same computations as before, we conclude that
$$
n_1(t)=\frac{2+4\lambda^2 t^2}{1+4\lambda^2 t^2}, \qquad n_2(t)=\frac{1+8\lambda^2 t^2}{1+4\lambda^2 t^2}.
$$
In this case, $n_1(t)$ decreases from 2 to 1, while $n_2(t)$ increases from 1 to 2, in agreement with our interpretation. Other choices of $\Psi(0)$ can be considered, and they all support our idea.

\subsection{Model 3: three agents and two levels}
The last model we want to consider in this section is a 3 { fermionic} agents system. This means that each agent has only two allowed levels. For that we consider the following three $8\times8$ matrices:
$$
b_1=\left(
\begin{array}{cccccccc}
 0 &  1 & 0 & 0 & 0 & 0 & 0 & 0\\
 0 &  0 & 0 & 0 & 0 & 0 & 0 & 0\\
 0 &  0 & 0 & 1 & 0 & 0 & 0 & 0\\
 0 &  0 & 0 & 0 & 0 & 0 & 0 & 0\\
 0 &  0 & 0 & 0 & 0 & 1 & 0 & 0\\
 0 &  0 & 0 & 0 & 0 & 0 & 0 & 0\\
 0 &  0 & 0 & 0 & 0 & 0 & 0 & 1\\
 0 &  0 & 0 & 0 & 0 & 0 & 0 & 0\\
\end{array}
\right),\quad b_2=\left(
\begin{array}{cccccccc}
0 &  0 & 1 & 0 & 0 & 0 & 0 & 0\\
0 &  0 & 0 & -1 & 0 & 0 & 0 & 0\\
0 &  0 & 0 & 0 & 0 & 0 & 0 & 0\\
0 &  0 & 0 & 0 & 0 & 0 & 0 & 0\\
0 &  0 & 0 & 0 & 0 & 0 & 1 & 0\\
0 &  0 & 0 & 0 & 0 & 0 & 0 & -1\\
0 &  0 & 0 & 0 & 0 & 0 & 0 & 0\\
0 &  0 & 0 & 0 & 0 & 0 & 0 & 0\\
\end{array}
\right),
$$
and
$$
b_3=\left(
\begin{array}{cccccccc}
0 &  0 & 0 & 0 & 1 & 0 & 0 & 0\\
0 &  0 & 0 & 0 & 0 & -1 & 0 & 0\\
0 &  0 & 0 & 0 & 0 & 0 & -1 & 0\\
0 &  0 & 0 & 0 & 0 & 0 & 0 & 1\\
0 &  0 & 0 & 0 & 0 & 0 & 0 & 0\\
0 &  0 & 0 & 0 & 0 & 0 & 0 & 0\\
0 &  0 & 0 & 0 & 0 & 0 & 0 & 0\\
0 &  0 & 0 & 0 & 0 & 0 & 0 & 0\\
\end{array}
\right),
$$
which satisfy the following CAR:
\be
\{b_k,b_j^\dagger\}=\delta_{k,j}\1, \qquad \qquad b_j^2=0,
\label{33}\en
$j,k=1,2,3$, where $\1$ is the identity operator on the Hilbert space of the system, $\Hil=\mathbb{C}^8$. Next we use these operators, and their adjoints, to construct an o.n. basis for $\Hil$. We start with $\varphi_{000}=(1\,\, 0\,\, 0\,\, 0\,\, 0\,\, 0\,\, 0\,\, 0 )^T$, (here $T$ is the transpose, as before). It is clear that $b_j\varphi_{000}=0$, $j=1,2,3$. Then we introduce
$$
\varphi_{100}=b_1^\dagger\varphi_{000}, \quad \varphi_{010}=b_2^\dagger\varphi_{000}, \quad \varphi_{001}=b_3^\dagger\varphi_{000}, \quad \varphi_{110}=b_1^\dagger b_2^\dagger\varphi_{000}, 
$$
$$
\varphi_{101}=b_1^\dagger b_3^\dagger\varphi_{000}, \quad\varphi_{011}=b_2^\dagger b_3^\dagger\varphi_{000},\quad \varphi_{111}=b_1^\dagger b_2^\dagger b_3^\dagger\varphi_{000},
$$
The set $\F_\varphi=\{\varphi_{ijk},\,i,j,k=0,1\}$ is an o.n. basis of $\Hil$.
We now consider the dynamics as driven by two  different Hamiltonians, $H_1=b_1^\dagger(\lambda b_2+\mu b_3)$ and $H_2=\lambda b_1^\dagger b_2+\mu b_2^\dagger b_3$, with $\lambda$ and $\mu$ positive quantities. The meaning of $H_1$  is easily understood: we claim that the action of $H_1$ increases $n_1(t)$ while decreasing both $n_2(t)$ and $n_3(t)$, where $N_j=b_j^\dagger b_j$ and $n_j(t)$ is found as in (\ref{32}). The effect of $H_2$ is less evident: it is clear that its action should lower $n_3(t)$ and increase $n_1(t)$, but not much can be said a priori on $n_2(t)$, since $H_2$ has two competing terms. However, we can imagine that the asymptotic value of $n_2(t)$ will be related to the relative magnitude of $\lambda$ and $\mu$. 

The matrix form of $H_1$ and of $U_1(t)=e^{-iH_1t}$ is the following:
$$
H_1=\left(
\begin{array}{cccccccc}
0 &  0 & 0 & 0 & 0 & 0 & 0 & 0\\
0 &  0 & \lambda & 0 & \mu & 0 & 0 & 0\\
0 &  0 & 0 & 0 & 0 & 0 & 0 & 0\\
0 &  0 & 0 & 0 & 0 & 0 & -\mu & 0\\
0 &  0 & 0 & 0 & 0 & 0 & 0 & 0\\
0 &  0 & 0 & 0 & 0 & 0 & \lambda & 0\\
0 &  0 & 0 & 0 & 0 & 0 & 0 & 0\\
0 &  0 & 0 & 0 & 0 & 0 & 0 & 0\\
\end{array}
\right),\quad U_1(t)=\left(
\begin{array}{cccccccc}
1 &  0 & 0 & 0 & 0 & 0 & 0 & 0\\
0 &  1 &-i \lambda t & 0 & -i\mu t & 0 & 0 & 0\\
0 &  0 & 1 & 0 & 0 & 0 & 0 & 0\\
0 &  0 & 0 & 1 & 0 & 0 & i\mu t & 0\\
0 &  0 & 0 & 0 & 1 & 0 & 0 & 0\\
0 &  0 & 0 & 0 & 0 & 1 & -i\lambda t & 0\\
0 &  0 & 0 & 0 & 0 & 0 & 1 & 0\\
0 &  0 & 0 & 0 & 0 & 0 & 0 & 1\\
\end{array}
\right).
$$
Let us put $\Psi(0)=\varphi_{011}$. This vector corresponds to $n_1(0)=0$ and $n_2(0)=n_3(0)=1$. If we now use formula (\ref{32}) we find
$$
n_1(t)=\frac{(\mu^2+\lambda^2)t^2}{1+(\mu^2+\lambda^2)t^2}, \quad n_2(t)=\frac{1+\mu^2t^2}{1+(\mu^2+\lambda^2)t^2}, \quad n_3(t)=\frac{1+\lambda^2t^2}{1+(\mu^2+\lambda^2)t^2},   
$$
which exhibit the desired behaviour: $n_1(t)$ increases from 0 to 1, $n_2(t)$ decreases from 1 to $\frac{\mu^2}{\mu^2+\lambda^2}$ and $n_3(t)$ decreases from 1 to $\frac{\lambda^2}{\mu^2+\lambda^2}$. The sum of $n_1(t)+n_2(t)+n_3(t)$ is always equal to 2. Hence, also in presence of a non self-adjoint Hamiltonian, the sum of the number operators can stay constant in time: $n_1(t)+n_2(t)+n_3(t)=2=n_1(0)+n_2(0)+n_3(0)$, which is an interesting feature of the framework, especially if can be generalized to other systems. 

If we start with a vector $\Psi(0)=\varphi_{010}$, repeating the same computations we get
$$
n_1(t)=\frac{\lambda^2t^2}{1+\lambda^2t^2}, \quad n_2(t)=\frac{1}{1+\lambda^2t^2}, \quad n_3(t)=0,
$$
which are again in agreement with the fact that $n_1(0)=n_3(0)=0$ and $n_2(0)=1$, and with the property of $H_1$ to destroy a state with $n_3(0)=0$, because of $b_3$, to increase $n_1(t)$ and to decrease $n_2(t)$. Similar considerations can be repeated with other choices of $\Psi(0)$. In particular, for some of them ($\varphi_{000}$, $\varphi_{110}$ or $\varphi_{101}$), the action  of $U_1(t)$ is trivial: nothing change.

Let us now see what happens if we use $H_2$ rather than $H_1$. In this case we have
$$
H_2=\left(
\begin{array}{cccccccc}
0 &  0 & 0 & 0 & 0 & 0 & 0 & 0\\
0 &  0 & \lambda & 0 & 0 & 0 & 0 & 0\\
0 &  0 & 0 & 0 & \mu & 0 & 0 & 0\\
0 &  0 & 0 & 0 & 0 & \mu & 0 & 0\\
0 &  0 & 0 & 0 & 0 & 0 & 0 & 0\\
0 &  0 & 0 & 0 & 0 & 0 & \lambda & 0\\
0 &  0 & 0 & 0 & 0 & 0 & 0 & 0\\
0 &  0 & 0 & 0 & 0 & 0 & 0 & 0\\
\end{array}
\right),\quad U_2(t)=\left(
\begin{array}{cccccccc}
1 &  0 & 0 & 0 & 0 & 0 & 0 & 0\\
0 &  1 &-i \lambda t & 0 & -\frac{\lambda\mu t^2}{2} & 0 & 0 & 0\\
0 &  0 & 1 & 0 & -i\mu t & 0 & 0 & 0\\
0 &  0 & 0 & 1 & 0 & -i\mu t & -\frac{\lambda\mu t^2}{2} & 0\\
0 &  0 & 0 & 0 & 1 & 0 & 0 & 0\\
0 &  0 & 0 & 0 & 0 & 1 & -i\lambda t & 0\\
0 &  0 & 0 & 0 & 0 & 0 & 1 & 0\\
0 &  0 & 0 & 0 & 0 & 0 & 0 & 1\\
\end{array}
\right).
$$
It is easy to check that, if $\Psi(0)=\varphi_{000}, \varphi_{100}, \varphi_{110}$ or  $\Psi(0)=\varphi_{111}$, then $n_j(t)=n_j(0)$ for all $t\geq0$ and $j=1,2,3$. This is because $\Psi(t)=U_2(t)\Psi(0)=\Psi(0)$ with each one of these choices. This, in turns, is in agreement with our interpretation of the model: in fact, $H_2$ destroys all these initial states. Let us now take an initial state which is not annihilated by $H_2$: $\Psi(0)=\varphi_{010}$. With this choice the usual computations give
$$
n_1(t)=\frac{\lambda^2t^2}{1+\lambda^2t^2}, \quad n_2(t)=\frac{1}{1+\lambda^2t^2}, \quad n_3(t)=0.
$$
We see that no dependence on $\mu$ appears in these formulas: in agreement with our interpretation of $H_2$, starting from $\varphi_{010}$, the term $\mu b_2^\dagger b_3$ in $H_2$ does not contribute, since it destroys $\varphi_{010}$, while the term $\lambda b_1^\dagger b_1$ is responsible of the dynamics above: $n_1(t)$ increases from 0 to 1, while $n_2(t)$ decreases from 1 to 0. $n_3(t)$ stays constantly equal to 0. Similar conclusions can be deduced if we fix, for instance, $\Psi(0)=\varphi_{001}$, with obvious replacements. More interesting is the situation if we consider $\Psi(0)=\varphi_{001}$, since both terms in $H_2$ act non trivially. In this case we find that
$$
n_1(t)=\frac{\lambda^2\mu^2t^4/4}{1+\mu^2t^2+\lambda^2\mu^2t^4/4}, \quad n_2(t)=\frac{\mu^2t^2}{1+\mu^2t^2+\lambda^2\mu^2t^4/4},  \quad n_3(t)=\frac{1}{1+\mu^2t^2+\lambda^2\mu^2t^4/4}.
$$
We see that both $\mu$ and $\lambda$ appear in these formulas. In particular we observe that $n_1(t)$ increases from 0 to 1, while $n_3(t)$ decreases from 1 to 0, as they should because of the form of $H_2$. As for $n_2(t)$, $H_2$ contains two competing terms, and in fact $n_2(t)$ increases from zero to its maximum value, $n_{2,max}=\frac{\mu}{\mu+\lambda}$, and then it decreases back to 0. Also, we observe that $n_{2,max}\simeq 1$ if $\lambda\ll \mu$, while $n_{2,max}\simeq 0$ if $\lambda\gg \mu$. This is because, if $\lambda\ll \mu$, $H_2\simeq \mu b_2^\dagger b_3$, so that $n_2(t)$ tends to increases to its maximum value, while if $\lambda\gg \mu$, $H_2\simeq \lambda b_1^\dagger b_2$, so that $n_2(t)$ does not change much from its original value, which is zero.

\section{An application to information dynamics}\label{sect4}

\begin{figure}[ht]
	\begin{center}
		\includegraphics[width=0.75\textwidth]{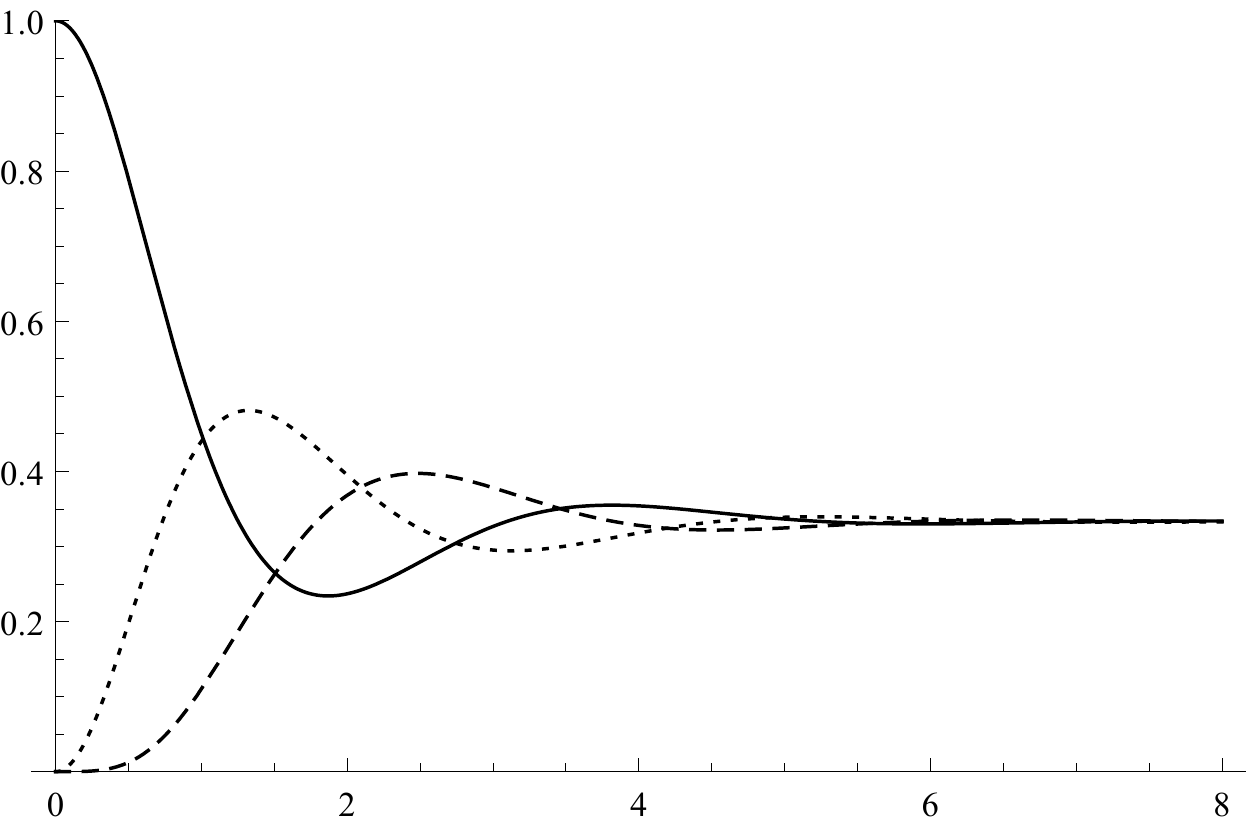}\hspace{8mm} %
	\end{center}
	\caption{{$n_1(t)$ (continuos line), $n_2(t)$ (dotted line), and $n_3(t)$ (dashed line) for $\Psi(0)=\varphi_{100}$.}}
	\label{fig1}
\end{figure}

The analysis carried out in Section \ref{sect3} allows us to conclude that the use of non self-adjoint Hamiltonians of the kind discussed so far is relevant, other than useful, if we are interested in describing {\em fluxes of quanta}\footnote{A flux of quanta is what we observe when the mean value of one number operator, $N_j$, decreases while the mean value of $N_k$, $k\neq j$, increases: there is a flux of quanta from agent $j$ to agent $k$.} going in one specific direction. This is possible, however, only if we use formula (\ref{32}) as the definition of the classical counterpart of the time evolution of the observable $X$ of the system. This is in agreement with the results in \cite{baggarg2}, where a similar strategy was already efficiently used in the description of cells proliferation. In this section we will consider an application of formula (\ref{32}), and of non self-adjoint Hamiltonians, to a simple model of information dynamics. We imagine a system $\Sc$ made of three agents, $\tau_1$, $\tau_2$ and $\tau_3$, which exchange some kind of (binary) information among them following two different schemes. The first case is ruled by the Hamiltonian
$$
H_a=b_2^\dagger b_1+b_3^\dagger b_2+b_1^\dagger b_3,
$$
which describes a one-directional flow of what we call {\em information} from $\tau_1$ to $\tau_2$, from $\tau_2$ to $\tau_3$ and then from $\tau_3$ back to $\tau_1$. Here the $b_j$'s are the same fermionic ladder operators satisfying (\ref{33}). If we take $\Psi(0)=\varphi_{100}$, it is possible to check that $\|\Psi(t)\|^2=\frac{1}{3}(1+2\cosh(\sqrt{3}\,t))$ and, for instance
$$
n_1(t)=\frac{1}{3(1+2\cosh(\sqrt{3}\,t))}\left(3+4\cosh\left(\frac{\sqrt{3}\,t}{2}\right)\cos\left(\frac{3t}{2}\right)+2\cosh(\sqrt{3}\,t)\right).
$$
Similar formulas are deduced for $n_2(t)$ and $n_3(t)$. In Figure \ref{fig1} we plot these functions.

We see that they all tend to a common asymptotic value, $n_j(\infty)\simeq\frac{1}{3}$, which is already reached for relatively small values of $t$, $t\simeq 7$ (in our units). The same value is reached if we consider different initial conditions, like $\Psi(0)=\varphi_{010}$ or $\Psi(0)=\varphi_{001}$. Analogously, if $\Psi(0)=\varphi_{110}, \varphi_{011}$ or  $\Psi(0)=\varphi_{101}$, we find $n_j(\infty)\simeq\frac{2}{3}$ while, if $\Psi(0)=\varphi_{000}$ or $\Psi(0)= \varphi_{111}$, $n_j(t)=n_j(0)$ for $j=1,2,3$. The conclusion is the following: $H_a$ describes a sort of {\em homogenization } of $\Sc$: the information, independently of how it was originally distributed, is spread uniformly among the three agents, which after some time are equally informed. Of course, this result implies that the dynamics produced by $H_a$, and possibly the dynamics deduced from all the Hamiltonians considered so far in this paper, is not reversible: once we arrive to the stationary state $n_j(\infty)\simeq\frac{1}{3}$, for instance, there is no way to understand if our original state was $\varphi_{100}$, $\varphi_{010}$ or $\varphi_{001}$, since they all produce the same $n_j(\infty)$. This irreversibility is not surprising, since $H_a\neq H_a^\dagger$.

\vspace{2mm}

\begin{figure}[h]
	\begin{center}
		\includegraphics[width=0.47\textwidth]{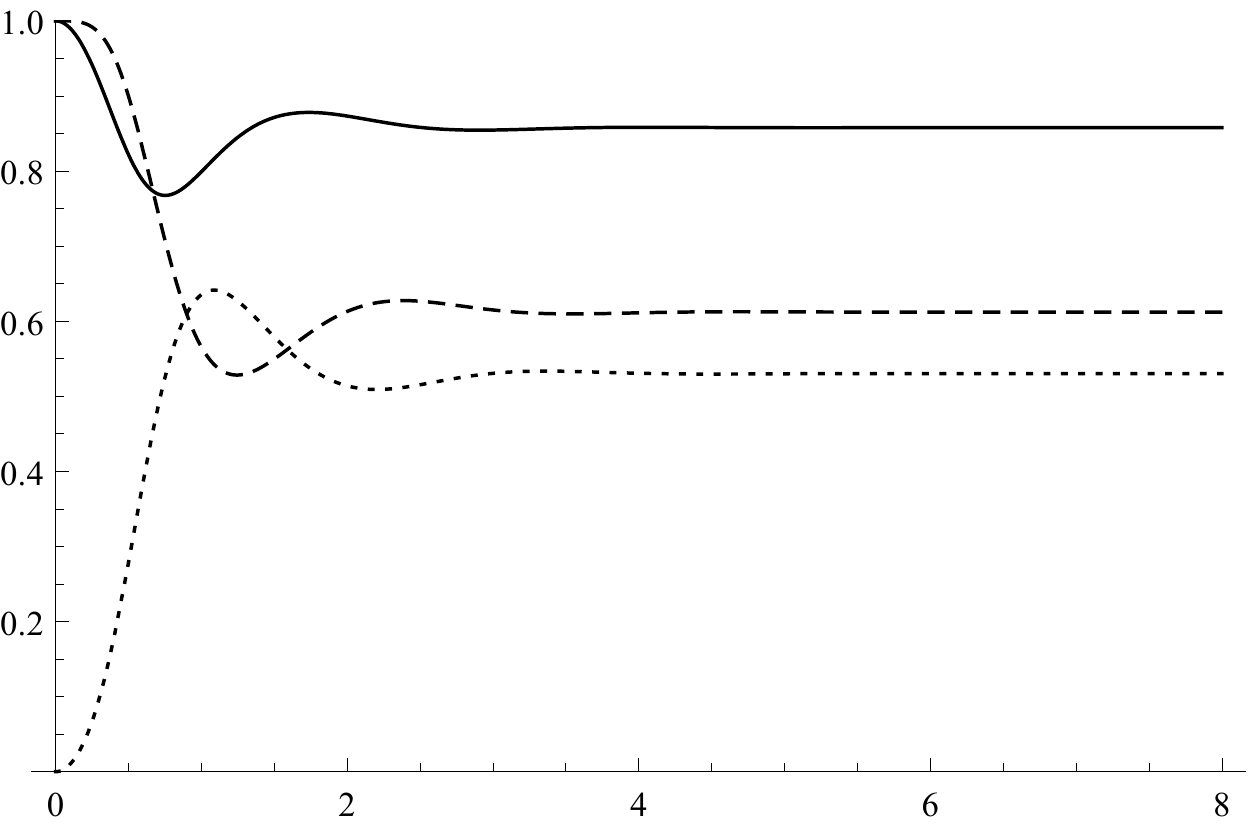}\hspace{8mm}
		\includegraphics[width=0.47\textwidth] {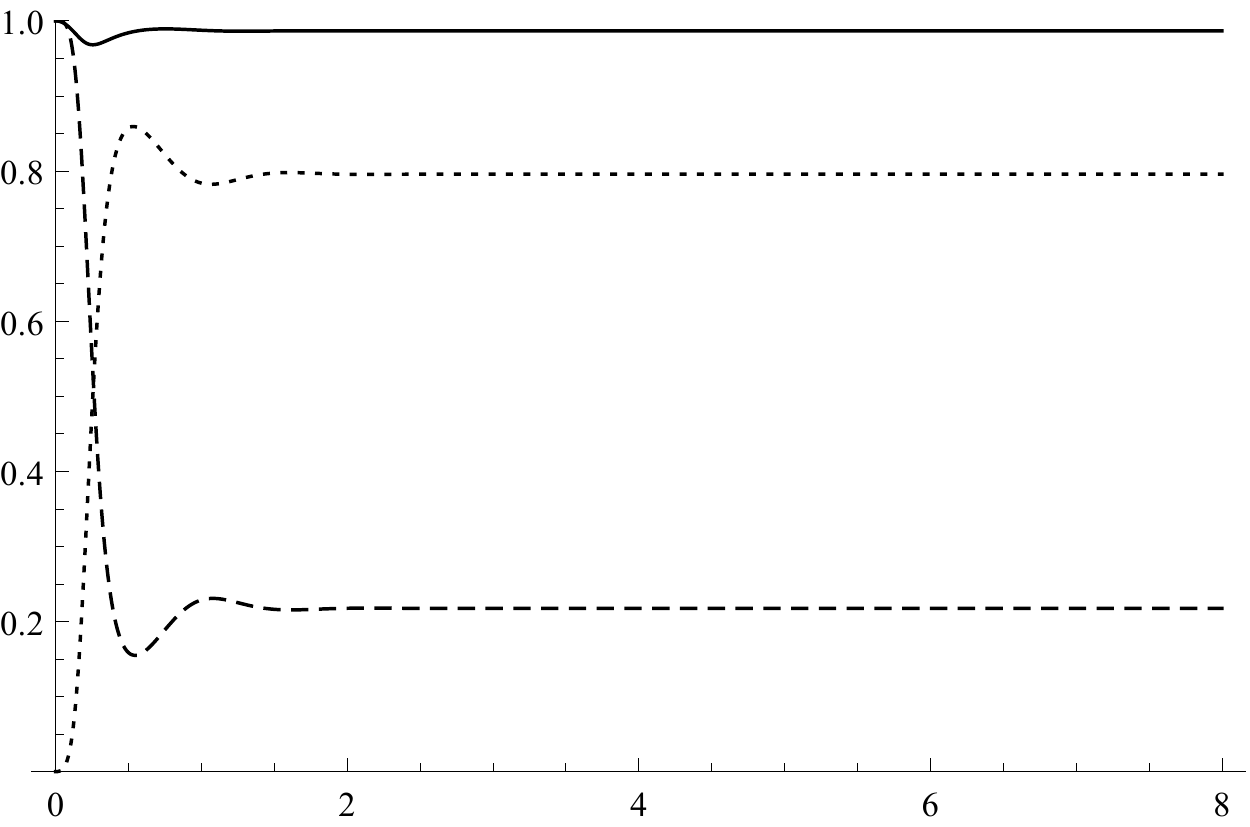}\hfill\\
		\includegraphics[width=0.47\textwidth] {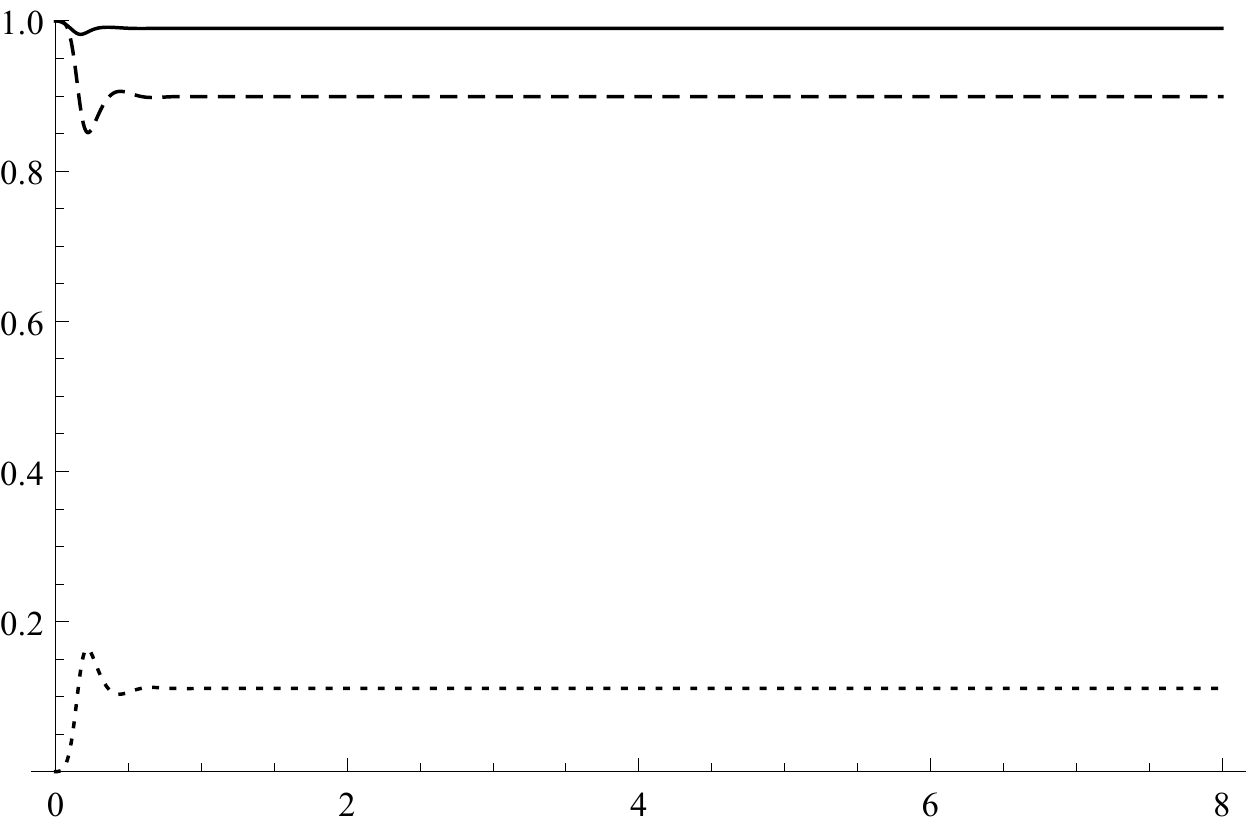}\\
		\caption{\label{fig2}\footnotesize $n_1(t)$ (continuous line), $n_2(t)$ (dotted line) and $n_3(t)$ (dashed line) for $\Psi(0)=\varphi_{101}$ and $(\lambda_1,\lambda_2,\lambda_3)=(1,2,3)$, upper left (a), $(\lambda_1,\lambda_2,\lambda_3)=(1,2,30)$, upper right (b), and $(\lambda_1,\lambda_2,\lambda_3)=(1,28,30)$, down (c).}
	\end{center}
\end{figure}

Of course, we do not expect that each Hamiltonian gives rise to such an homogenization. This is what we will show now using the following operator
$$
H_b=\lambda_1b_2^\dagger b_1+\lambda_2b_3^\dagger b_2+\lambda_3b_1^\dagger b_3,
$$
which is an anisotropic version of $H_a$, where the strength of the various interactions, fixed by the various $\lambda_j$'s,  can now be different. The computations follow the same steps as above, and will not be repeated. The plots in Figure \ref{fig2} show the time evolution of $n_1(t)$, $n_2(t)$ and $n_3(t)$ for three different choices of the $\lambda_j$'s and for $\Psi(0)=\varphi_{101}$ (i.e., for $n_1(0)=n_3(0)=1$ and $n_2(0)=0$). In particular, in Figure \ref{fig2} (a) we have taken $(\lambda_1,\lambda_2,\lambda_3)=(1,2,3)$,  in Figure \ref{fig2} (b) we have  $(\lambda_1,\lambda_2,\lambda_3)=(1,2,30)$, and in Figure \ref{fig2} (c) we have $(\lambda_1,\lambda_2,\lambda_3)=(1,28,30)$. This is to consider the case of slightly or quite different $\lambda_j$'s.    

In all cases we see that an asymptotic value is reached by each $n_j(t)$, and the speed increases for larger values of the $\lambda_j$'s. Also, when the numerical values of $\lambda_j$ increase, the functions $n_j(t)$ change less and less during the time evolution. It is evident that there is no homogenization here. This is easily understood, since the three $\lambda_j$'s are all different. Hence, $H_b$ produces differences in the information of the agents, during the time evolution and in their asymptotic values. For instance, when   $(\lambda_1,\lambda_2,\lambda_3)=(1,2,30)$, we see that even if $n_2(0)<n_3(0)$, after a short time the inequality is reversed, and stay reversed for the rest of the time evolution: the originally better informed agent becomes soon worst informed, in fact. Similar conclusions can be deduced for other choices of $\Psi(0)$. In particular, in all cases the system reach an equilibrium after some time, and this time is smaller when some of the $\lambda_j$'s is large.

\section{Conclusions}\label{sect5}

In this paper we have discussed the role of non self-adjoint Hamiltonians in the analysis of macroscopic systems, when  uni-directional fluxes of some {\em quanta-like} quantities is expected. We have seen that, extending the general procedure proposed in \cite{bagbook,bagbook2}, it is possible to use some $H\neq H^\dagger$ to describe quantities which goes from one agent to another, and do not go back. The minor price to pay is that the state of the system looses normalization when $t$ grows, and it must be restored by hand as in (\ref{32}), which becomes the key equation for us. As we have already noticed, this equation returns the standard formula for mean values of operators when $H=H^\dagger$. 

We have applied our strategy to a series of preliminary simple closed systems, i.e. to system which, contrarily to what is done in other papers, \cite{bag5,baggarg,baghavkhr}, do not interact with any reservoir. In this way, we have checked the self-consistency of the technique. Afterwards we have applied the same strategy to a simple model of information exchanged among agents, and we have found mechanisms which homogenize the agents, and mechanisms which keep them differently informed. The analysis here is further confirmed by a recent application to tumour growth, \cite{baggarg2}. More applications will be considered soon.

\section*{Acknowledgements}

This work was partially supported by the University of Palermo and by the Gruppo Nazionale di Fisica Matematica of Indam.

\end{document}